\documentclass[aps,twocolumn,floatfix]{revtex4}
\usepackage{epsfig}
\usepackage{graphicx}
\usepackage{dcolumn}
\usepackage{natbib}
\usepackage{amssymb}

\def\dj{\hbox{d\kern-0,347em \vrule width0,3em height1,252ex
depth-1,21ex \kern0,051em}}
 
\begin{document}

\title{Statistical properties of fracture in a random spring model}

\author{Phani Kumar V.V. Nukala}
\affiliation{Computer Science and Mathematics Division, 
Oak Ridge National Laboratory, Oak Ridge, TN 37831-6359, USA}
\author{Stefano Zapperi}
\affiliation{INFM UdR Roma 1 and SMC, Dipartimento di Fisica,
Universit\`a "La Sapienza", P.le A. Moro 2, 00185 Roma, Italy}
\author{Sr{\dj}an \v{S}imunovi\'{c}}
\affiliation{Computer Science and Mathematics Division, 
Oak Ridge National Laboratory, Oak Ridge, TN~37831-6359, USA}
 
\begin{abstract}
Using large scale numerical simulations we analyze the statistical
properties of fracture in the two dimensional random spring model and
compare it with its scalar counterpart: the random fuse model.  We
first consider the process of crack localization measuring the
evolution of damage as the external load is raised. We find that, as
in the fuse model, damage is initially uniform and localizes at peak
load. Scaling laws for the damage density, fracture strength and
avalanche distributions follow with slight variations the behavior
observed in the random fuse model.  We thus conclude that scalar
models provide a faithful representation of the fracture properties of
disordered systems.
\end{abstract}
\maketitle

\section{introduction}

The statistical properties of fracture in disordered media represent
an intriguing theoretical problem with important practical
applications \cite{breakdown}. The presence of disorder naturally leads 
to statistical distributions of failure stresses, accumulated damage,
acoustic activity, crack shapes and so on.  The application of a
standard continuum descriprion based on elastic equation cannot
capture the effect of fluctuations and hence the effect of disorder has to be considered 
explicitly.  A well established way to deal with this problem relies on
lattice models, in which the medium is described by a discrete set of elastic
bonds with randomly distributed failure thresholds
\cite{breakdown}.  In the simplest approximation of a
scalar displacement, one recovers the random fuse model (RFM) where a
lattice of fuses with random thresholds are subject to an increasing
external current \cite{deArcangelis85}.

The RFM has been extensively investigated in the last twenty years,
mainly using numerical simulations
\cite{breakdown,deArcangelis85,kahng88,deArcangelis89,delaplace,zrvs,hansen001,nukalajpamg}. The
type of behavior at macroscopic fracture is significantly influenced
by the amount of disorder \cite{kahng88}.  When the disorder is
narrowly distributed, materials breakdown without significant
precursors. As the disorder increases, substantial damage is
accumulated prior to failure and the dynamics resembles percolation
\cite{sornette}.  Indeed, in the limit of infinite disorder, the
damage accumulation process can exactly be mapped onto a percolation
problem \cite{guyon88}. It has been suggested that for strong, but
finite, disorder fracture should be interpreted as a first order
transition near a spinodal point \cite{zrvs}.  In addition, the
fracture of the RFM is preceded by avalanches of failure events
\cite{zrvs,hansen,alava,ZAP-05}. These are reminiscent of the acoustic
emission activity observed in experiments and their distribution
follows a power law.  Finally, the RFM has also been used to compute
the fracture strength distribution and the related size effects 
\cite{duxbury86,duxbury87,duxbury88,beale88,nukalaepjb}.

Modeling the elastic medium using the RFM 
introduces drastic approximations in terms of the discretization
process, quasistatic dynamics and the scalar nature
of the interactions. It is thus important to clarify if the 
observations made in the RFM carry over to more complex and realistic
situations. In this paper, we address the problem of the scalar
(electric) interactions of the RFM, by comparing it with a tensorial
central force model, the random spring model (RSM) \cite{sahimi86}. 
The model is a tensorial counterpart of the RFM: it has quasistatic dynamics,
random thresholds, but fuses and currents are replaced by 
elastic springs and forces. Dynamic effects have been instead
considered in Refs. \cite{minozzi,politi}. 

After discussing the model in Sec. II, we 
consider the typical statistical measures performed using the
RSM: damage localization and average damage profiles
are reported in Sec. III, while mean damage scaling and 
damage distributions are discussed in Sec. IV. and Sec. V respectively. 
In Sec. VI and VII we discuss the fracture strenght distribution
and the size effect on the mean strength. The avalanche behavior
is analyzed in Sec. VII and a summary is reported in Sec VIII.
We have not analyzed the roughness of the final crack since
in several instances the spring networks fail because of 
loss of rigidity.

\section{The random spring model}

In the RSM, the lattice is initially fully intact
with bonds having the same stiffness, but the bond breaking
thresholds, $t$, are randomly distributed based on a thresholds
probability distribution, $p(t)$. 
The bond breaks irreversibly, whenever the force in the spring exceeds the
breaking threshold force value, $t$, of the spring. Periodic boundary
conditions are imposed in the horizontal direction and a constant unit displacement 
difference is applied between the top and the bottom of lattice system. 

\begin{figure}[hbtp]

\includegraphics[width=8cm]{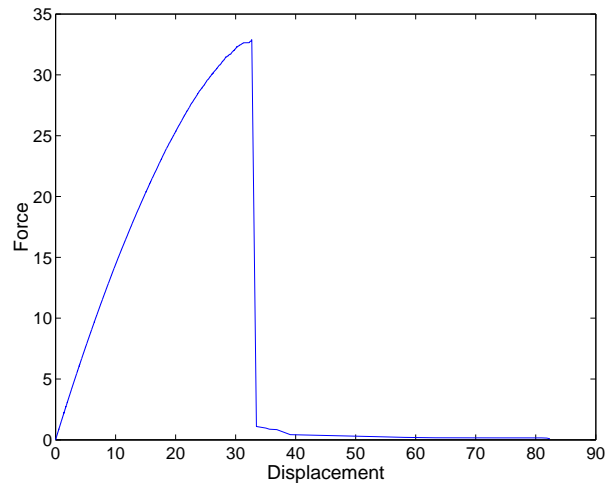}
\caption{Envelope of a typical force-displacement response obtained using the RSM.}
\label{fig:force_disp}
\end{figure}

Numerically, a unit displacement, $\Delta = 1$, is applied at the top
of the lattice system and the equilibrium equations are solved to
determine the force in each of the springs. Subsequently, for each
bond $j$, the ratio between the force $f_j$ and the breaking threshold
$t_j$ is evaluated, and the bond $j_c$ having the largest value,
$\mbox{max}_j \frac{f_j}{t_j}$, is irreversibly removed. The forces
are redistributed instantaneously after a bond is broken implying that
the stress relaxation in the lattice system is much faster than the
breaking of a bond. Each time a bond is broken, it is necessary to
re-equilibrate the lattice system in order to determine the subsequent
breaking of a bond.  The process of breaking of a bond, one at a time,
is repeated until the lattice system falls apart. For the RSM, we
consider a triangular lattice system network and a uniform probability
distribution for thresholds disorder, which is constant between 0 and
1. The diamond lattice (square lattice with bonds inclined at 45 degrees) 
spring system exhibits certain unstable modes
and hence is not considered. Figure \ref{fig:force_disp} presents the 
envelope of a typical force-displacement response obtained using the RSM. 
The peak load of the lattice system is defined as the maximum force 
of the force-displacement response. 

Numerical simulation of fracture using large lattice networks is often
hampered due to the high computational cost associated with solving a
new large set of linear equations every time a new lattice bond is
broken. In this study, we use the multiple-rank sparse Cholesky
factorization updating algorithm developed in Ref. \cite{nukalajpamg}
for simulating fracture using discrete lattice systems.  In comparison
with the Fourier accelerated iterative schemes used for modeling
lattice breakdown \cite{bat-98}, this algorithm significantly
reduced the computational time required for solving large lattice
systems. Using this numerical algorithm, we were able to investigate
damage evolution in large ($L=512$ for spring model) initially
fully intact discrete lattice systems.  However, due to insufficient
number of available sample configurations, in this paper, we consider
results up to $L=256$ for spring models.  For many lattice system
sizes, the number of sample configurations, $N_{config}$, used are
excessively large to reduce the statistical error in the numerical
results (see Table 1). In Table 1, the fraction of broken bonds (or
damage density) for each of the lattice system sizes is obtained by
dividing the number of broken bonds with the total number of bonds,
$N_{el}$, present in the fully intact lattice system. For triangular
lattice topology, $N_{el} = (3L+1) (L+1)$. The lattice system sizes
considered in this work are $L = \{8, 16, 24, 32, 64, 128, 256\}$.
However, since corrections to the scaling laws are strongest for small
lattice systems, in the following, we use lattice sizes $L \ge 16$ for
obtaining the scaling exponents.  Table 1 presents mean and standard
deviations in the broken bond density (fraction of broken bonds) at
the peak load and at failure for various triangular lattice system
sizes.

\section{Diffusive Damage and Localization}
Qualitatively, damage evolution as described by breaking of bonds is
controlled by two competing aspects: disorder and stress concentration
in the vicinity of crack tips.  In the case of strong disorder, bond
breaking events occur in an uncorrelated manner in the initial stages
of damage evolution and thus resemble percolation.  As the damage
starts to accumulate, some degree of correlation can be expected due
to the presence of stress concentration at the crack tips.  A natural
question to ask concerns the relevance of these correlations as
failure is approached. If correlations are irrelevant one should
observe percolation scaling up to failure, as in the case of infinite
disorder. On the other hand, in the weak disorder case, the current
enhancement at the crack tips is so strong that a spanning crack is
nucleated soon after a few bonds (or even a single bond) are broken
\cite{kahng88}.  The interesting situation corresponds to the diffuse
damage and localization regime, where a substantial amount of damage
is accumulated prior to failure. Figure ~\ref{fig:damage_evol} presents the 
snapshots of damage evolution in a typical RSM simulation of size $L = 256$.

In order to investigate the localization of damage prior to failure,
we divided the load-displacement response of a typical RSM simulation
into 12 segments, with six equal segments each before and after the
peak load. Figures \ref{fig:series_pre} and \ref{fig:series_post}
present the snapshots of damage profiles within each segment of
load-displacement curve of a typical simulation with uniform threshold
disorder for $L = 256$. Based on Figs.~\ref{fig:series_pre} and
\ref{fig:series_post}, it is clear that localization of damage occurs
in the RSM prior to failure even for strong but finite
disorder. In fact, the damage is diffusive in the initial stages of
loading and extends upto almost the peak load. Around the peak load,
the damage starts to localize and ultimately leads to failure, and
hence the final breakdown event is very different from the initial
precursors upto the peak load. Similar behavior is observed in the
random thresholds fuse model with both uniform and power law
thresholds distributions \cite{nukalajstat}.

\begin{figure*}[hbtp]
\caption{Snapshots of damage evolution in a typical simulation of size $L = 256$. Number of broken bonds at 
the peak load and at failure are 13864 and 16695, respectively. (1)-(9) represent the snapshots of damage
after $n_b$ bonds are broken. (1) $n_b = 5000$ (2) $n_b = 10000$ 
(3) $n_b = 12000$ (4) $n_b = 13000$ 
(5) $n_b = 14000$ (just after peak load) (6) $n_b = 15000$ (7) $n_b = 15500$ 
(8) $n_b = 16000$ (9) $n_b = 16500$ (close to failure)}
\label{fig:damage_evol}
\end{figure*}

\begin{figure}[hbtp]

\includegraphics[width=8cm]{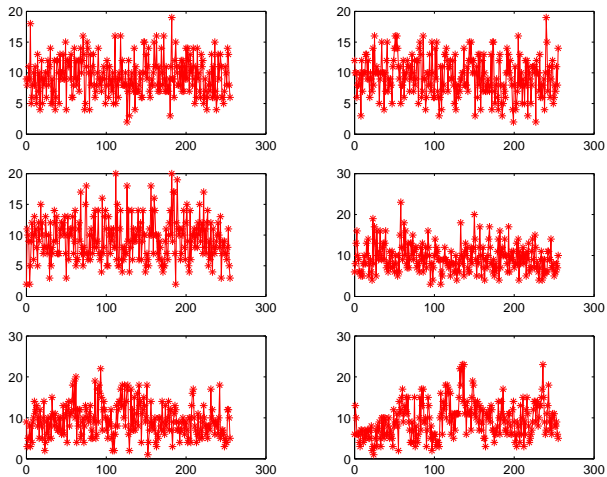}
\caption{Snapshots of pre-peak damage profiles of a typical RSM simulation with 
uniform threshold distribution on a triangular lattice of size $L = 256$. 
The damage is uniform in the pre-peak regime.}
\label{fig:series_pre}
\end{figure}

\begin{figure}[hbtp]

\includegraphics[width=8cm]{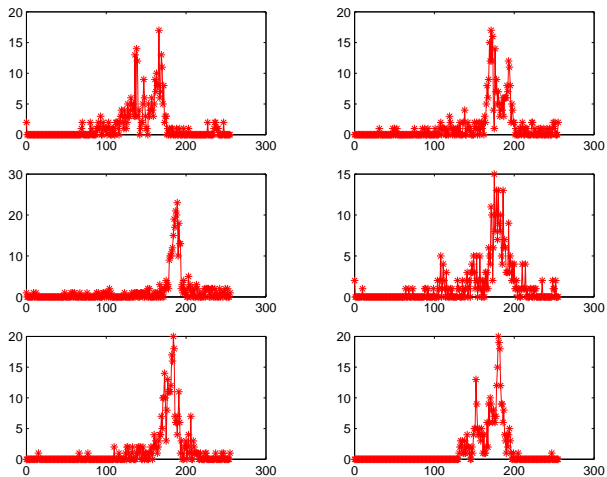}
\caption{Snapshots of post-peak damage profiles of a typical RSM simulation with 
uniform threshold distribution on a triangular lattice of size $L = 256$. 
The damage is clearly localized in the post-peak regime.}
\label{fig:series_post}
\end{figure}

In order to obtain a quantitative description of the damage
localization process it is necessary to average the damage profiles
over different realizations. Since the localization of damage can
occur anywhere along the $y$ direction of the lattice, a simple
averaging of the damage profiles would yield a flat profile
irrespective of the individual profile shapes in a single
realization. In this study, we average the damage profiles by first
shifting the damage profiles by the center of mass of the damage and
then averaging. Alternatively, one could average the magnitude of the
Fourier transforms of individual damage profiles thereby retaining the
frequency content of damage profiles. The Fourier method eliminates
any artificial biasing associated with the shifting of the individual
profiles in the real-space \cite{nukalajstat}.

\begin{figure}[hbtp]
\includegraphics[width=8cm]{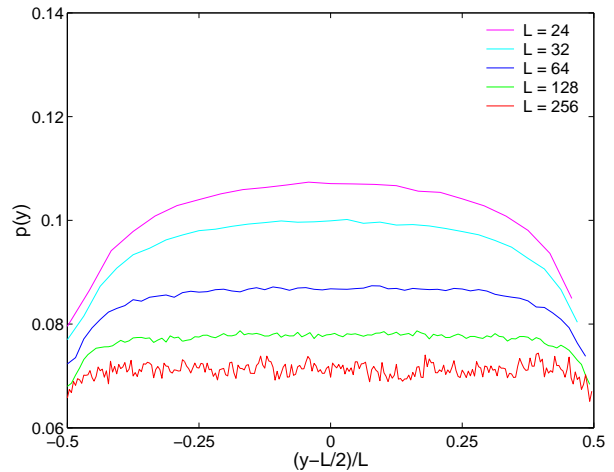}
\caption{Average damage profiles at peak load obtained by first centering the data
around the center of mass of the damage and then averging over different samples. For 
each of the samples, the damage profile is evaluated as $p(y) = \frac{n_b(y)}{(3 L + 1)}$, 
where $n_b(y)$ denotes the number of broken bonds in the $y^{th}$ section.}
\label{fig:loc2}
\end{figure}

\begin{figure}[hbtp]
\includegraphics[width=8cm]{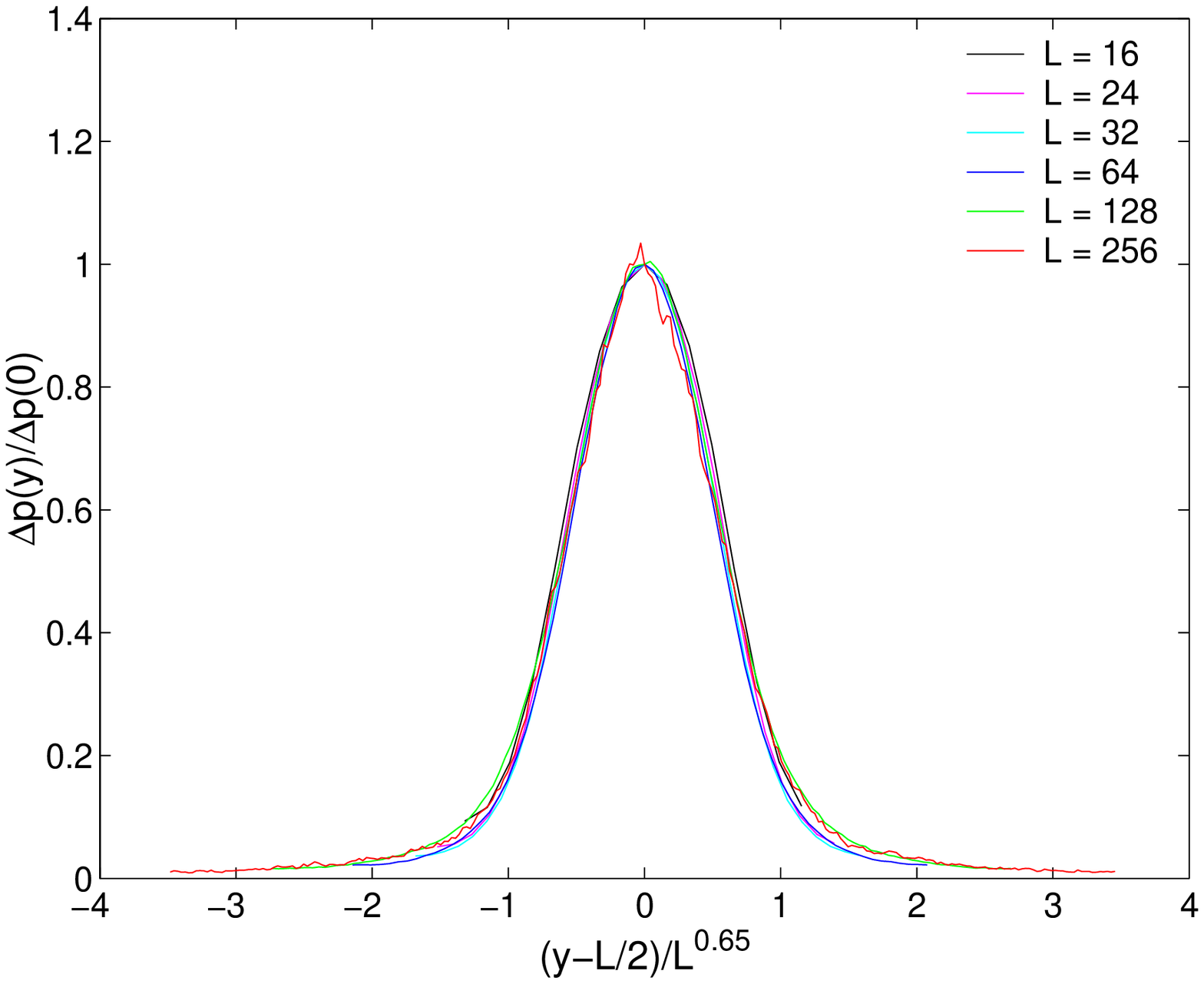}
\caption{Data collapse of the average profiles 
for the damage accumulated between peak load and failure using a power law scaling. 
We have considered the damage profiles for $L = \{16,24,32,64,128,256\}$ system sizes. 
The average has been performed after shifting by the center of mass. The profiles
show exponential tails.}
\label{fig:profCM_coll}
\end{figure}

Figure \ref{fig:loc2} presents the average damage profiles for the
damage accumulated up to the peak load by first shifting the damage
profiles by the center of mass of the damage and then averaging over
different samples. The results presented in Fig. \ref{fig:loc2}
indicate that although the average damage profiles at smaller lattice
system sizes are not completely flat, they flatten considerably as the
lattice system size is increased. We tend thus to attribute the
apparent profile to size effects. Indeed, for large system sizes
(e.g. $L = 128$ and $256$), the results clearly show that there is no
localization at the peak load. Consequently, the localization of
damage is mostly due to the damage accumulated between the peak load
and failure, i.e., the final catastrophic breakdown event. Figure
\ref{fig:profCM_coll} presents the data collapse of the average damage
profiles for the damage accumulated between the peak load and failure
using a power law scaling. A perfect collapse of the data is obtained
using the form
\begin{equation}
\langle \Delta p(y,L) \rangle/\langle \Delta p(0) \rangle= f(|y-L/2|/\xi),
\label{eq:prof2}
\end{equation}
where the damage peak scales as $\langle \Delta p(0) \rangle = L^{-0.37}$ and the 
localization length scales as $\xi \sim L^\alpha$, with $\alpha=0.65$ (see Fig.~\ref{fig:profCM_coll}). 
The profile shapes decay exponentially at large system sizes. 
We have also tried a simple linear scaling of the form 
$\langle \Delta p(y,L) \rangle/\langle \Delta p(0) \rangle = f((y-L/2)/L)$,
but the collapse of the data is not very good. The result for the fuse model is
similar: the profile also displays exponential tails and the exponent is found to be
 $\alpha= 0.8$.

\section{Scaling of damage density}
\begin{figure}[hbtp]
\centerline{\psfig{file=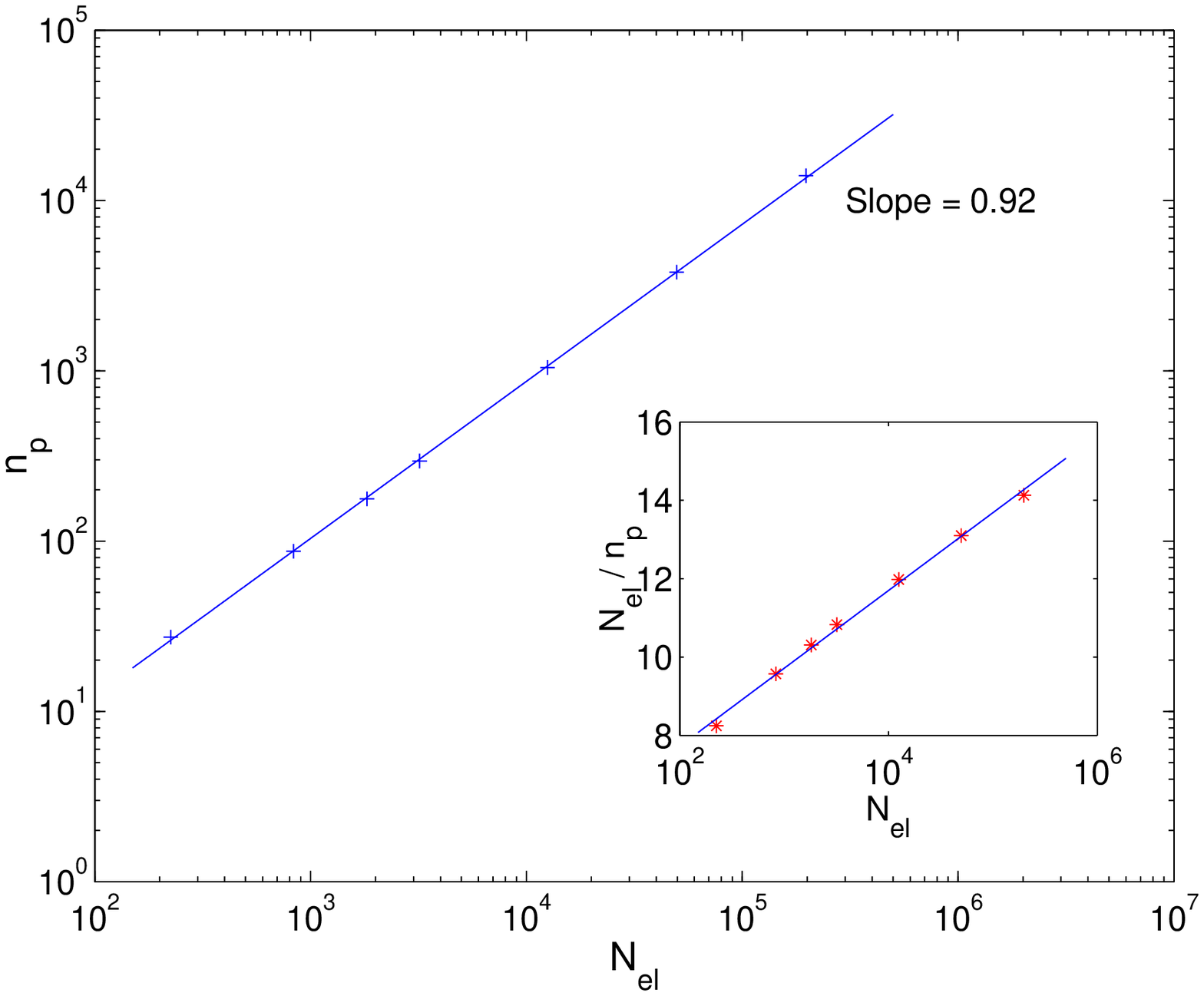,width=8cm,clip=!}}
\caption{Scaling of number of broken bonds at peak load for triangular random 
thresholds spring lattices. The scaling exponent $b=0.92$ is very close to 
the exponent obtained for random thresholds fuse network using 
triangular ($b=0.93$) and diamond ($b=0.91$) lattices. 
The difference could be attributed to small size effects. 
The number of broken bonds at peak load can also be
fit by a linear function times a logarithmic correction by 
plotting $n_p/N_{el}$ as a function of $N_{el}$ in a log-linear plot (inset).}
\label{fig:2}
\end{figure}

It has been noted in the previous section that the final breakdown event is very 
different from the initial precursors. Thus, we 
consider the scaling of the number of broken bonds at the peak load, $n_p$,
that excludes the last catastrophic event. In Fig.~\ref{fig:2} we plot
$n_p$ as a function of the lattice size $N_{el}$. The data displays a reasonable power law behavior
$n_p \sim N_{el}^b$, with $b=0.92$. The exponent $b=0.92$ is 
in close agreement with the value obtained for random thresholds fuse model 
using both triangular ($b=0.93$) and diamond ($b=0.91$) lattice topologies \cite{nukalajstat}. 
The difference between the RSM and RFM models is marginal and may be attributed to the 
results obtained from the smaller lattice sizes, where corrections to the fractal scaling 
may exist. However, we have noticed some systematic deviations from the scaling form $n_p \sim N_{el}^b$ 
by plotting $n_p/N_{el}^b$ vs $N_{el}$. Since the exponent 
$b$ is close to one, the data could be equally well fit by a linear law times a
logarithmic correction $n_p \simeq N_{el}/\log(N_{el})$ as suggested
in Ref.~\cite{mikko04} (see the inset of Fig.~\ref{fig:2}). Both these fits imply
that in the limit of large lattices the fraction of broken bonds prior
to fracture vanishes.

\section{Scaling of Damage Density Probability Distributions}
\begin{figure}[hbtp]
\centerline{\psfig{file=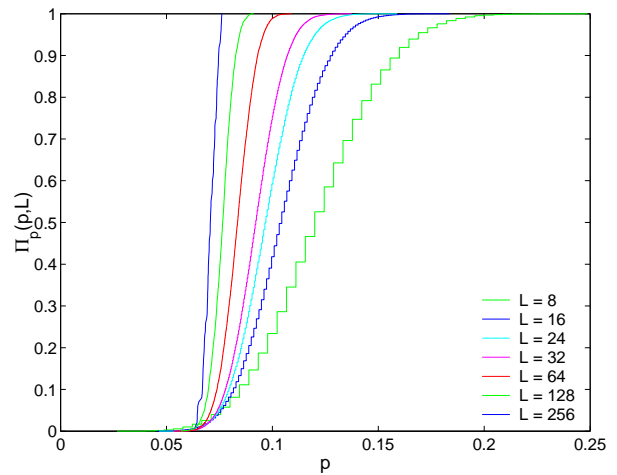,width=8cm,clip=!}}
\caption{The cumulative probability distribution for the fraction of broken
bonds at the peak load for triangular spring lattices of different system sizes.}
\label{fig:5}
\end{figure}

\begin{figure}[hbtp]
\centerline{\psfig{file=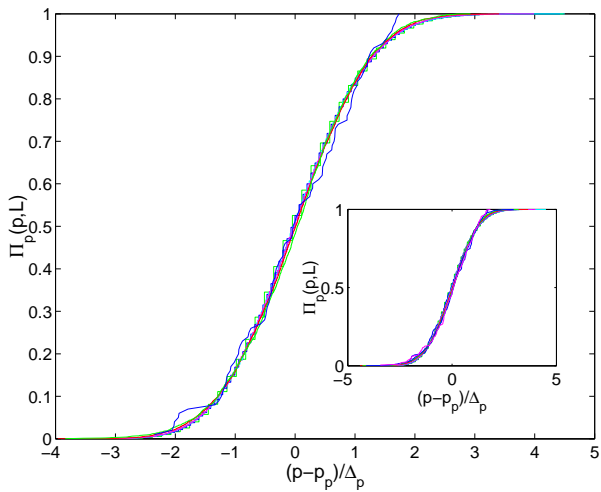,width=8cm,clip=!}}
\caption{The collapsed cumulative probability distribution for the fraction of 
broken bonds at the peak load in the random thresholds spring model (RSM) using 
triangular lattices of different system sizes ($L = 8,16,24,32,64,128,256$) 
with uniform disorder when plotted as a function of the reduced variable 
$\bar{p}_p=(p-p_p)/\Delta_p$. In the inset, a comparison 
between the cumulative probability distributions of the 
fraction of broken bonds at the peak load is presented for the RSM and RFM. 
For the RSM, triangular lattices of sizes ($L = 8,16,24,32,64,128,256$), and 
for the RFM, triangular lattices of sizes ($L = 16,24,32,64,128,256,512$) 
are plotted. In the RFM case, collapse of cumulative probability distributions 
at the peak load for different lattice topologies (triangular and diamond) and different 
disorder distributions (uniform and power law) is presented in Ref. \cite{nukalajstat}.}
\label{fig6}
\end{figure}

\begin{figure}[hbtp]
\centerline{\psfig{file=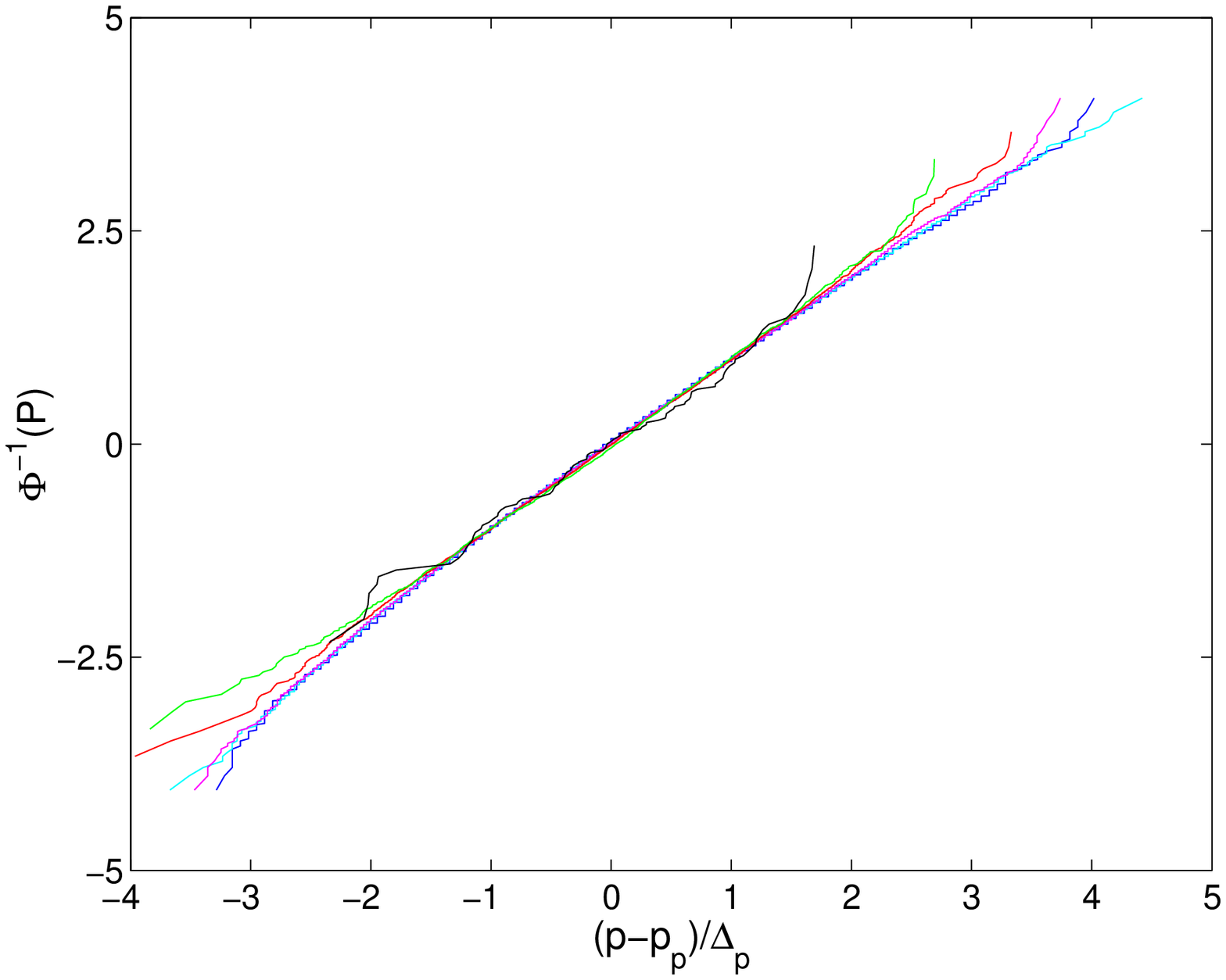,width=8cm,clip=!}}
\caption{Normal distribution fit for the cumulative probability distributions 
of the fraction of broken bonds at the peak load 
for triangular spring lattices of different system sizes $L$ = \{8, 16, 24, 32, 64, 128, 256\}.}
\label{fig9}
\end{figure}

Since the final breakdown event is very different from the initial precursors 
up to the peak load, in this section, we present the scaling of the 
cumulative probability distributions for the fraction of broken
bonds at the peak load. The cumulative probability distribution for 
the damage density at the peak load is defined as 
the probability $\Pi_{p} (p_b,L)$ that a system of size $L$ reaches peak load when the 
fraction of broken bonds equals $p_b = \frac{n_b}{N_{el}}$, where $n_b$ is the number of broken bonds. 
Figure~\ref{fig:5} presents the cumulative probability distribution for the damage density at the 
peak load in the random thresholds spring model for various system sizes $L$. 
By  simply plotting
the distribution in terms of $\bar{p}_p \equiv \frac{(n_b -
\mu_{n_p})}{\sigma_{n_p}} = \frac{(p_b - p_{p})}{\Delta_{p}}$, where
$\mu_{n_p}$ and $\sigma_{n_p}$ denote the mean and standard deviation
of the number of broken bonds at peak load, and $p_p$ and $\Delta_{p}$
denote the mean and standard deviation of fraction of broken bonds at
peak load (see Table 1), we obtain a very good collapse of the cumulative probability 
distribution of the damage density at the peak load. Fig.~\ref{fig6} shows that $\Pi_{p} (p,L)$ may be
expressed in a universal scaling form such that $\Pi_{p} (p,L) =
\Pi_{p} (\bar{p}_p)$ for different system sizes $L$.  A similar collapse
has been performed for the random thresholds fuse model in the Ref.~\cite{nukalajstat}. 
The inset in Fig.~\ref{fig6} presents the comparison of the cumulative damage density probability distributions 
in the random thresholds spring and fuse models. 
The excellent collapse of the data in the inset of Fig. \ref{fig6} suggests that the cumulative 
probability distribution for the damage density at the peak load, $\Pi_{p} (p_b,L) = \Pi_{p} (\bar{p}_p)$, 
may be universal. Finally, the collapse of the data
in Fig. \ref{fig9} indicates that a Gaussian distribution adequately
describes $\Pi_p$.

In Ref.~\cite{nukalajstat}, we have also checked that the distributions at failure 
in the random thresholds fuse model obey essentially the same laws, i.e., $\Pi (\bar{p}) = \Pi_{f} (\bar{p}_f)
= \Pi_{p} (\bar{p}_p)$, where $\Pi_{f} (\bar{p}_f)$ is the probability
that a system of size $L$ fails when the fraction of
broken bonds equals $p_b$, and $\bar{p}_f$ is the corresponding
reduced variable at failure. However, in the RSM, although a reasonable collapse of 
the cumulative probability distribution of damage density at failure 
can be obtained, the cumulative distributions of damage density at peak load and at failure 
appear to be different. In particular, the distribution, $\Pi_{f} (\bar{p}_f)$, at failure is not 
adequately described by a Gaussian distribution. The inadequacy of a Gaussian distribution 
in the post-peak regime may indicate the presence of a relatively stronger localization 
in the RSM compared with the RFM.

\section{Universality of fracture strength distribution}
In this study, we start the numerical simulation with a fully intact lattice system. The 
fracture strength of such a system is defined as the stress corresponding to the 
peak load of the lattice system response. 

\begin{figure}[hbtp]

\includegraphics[width=8cm]{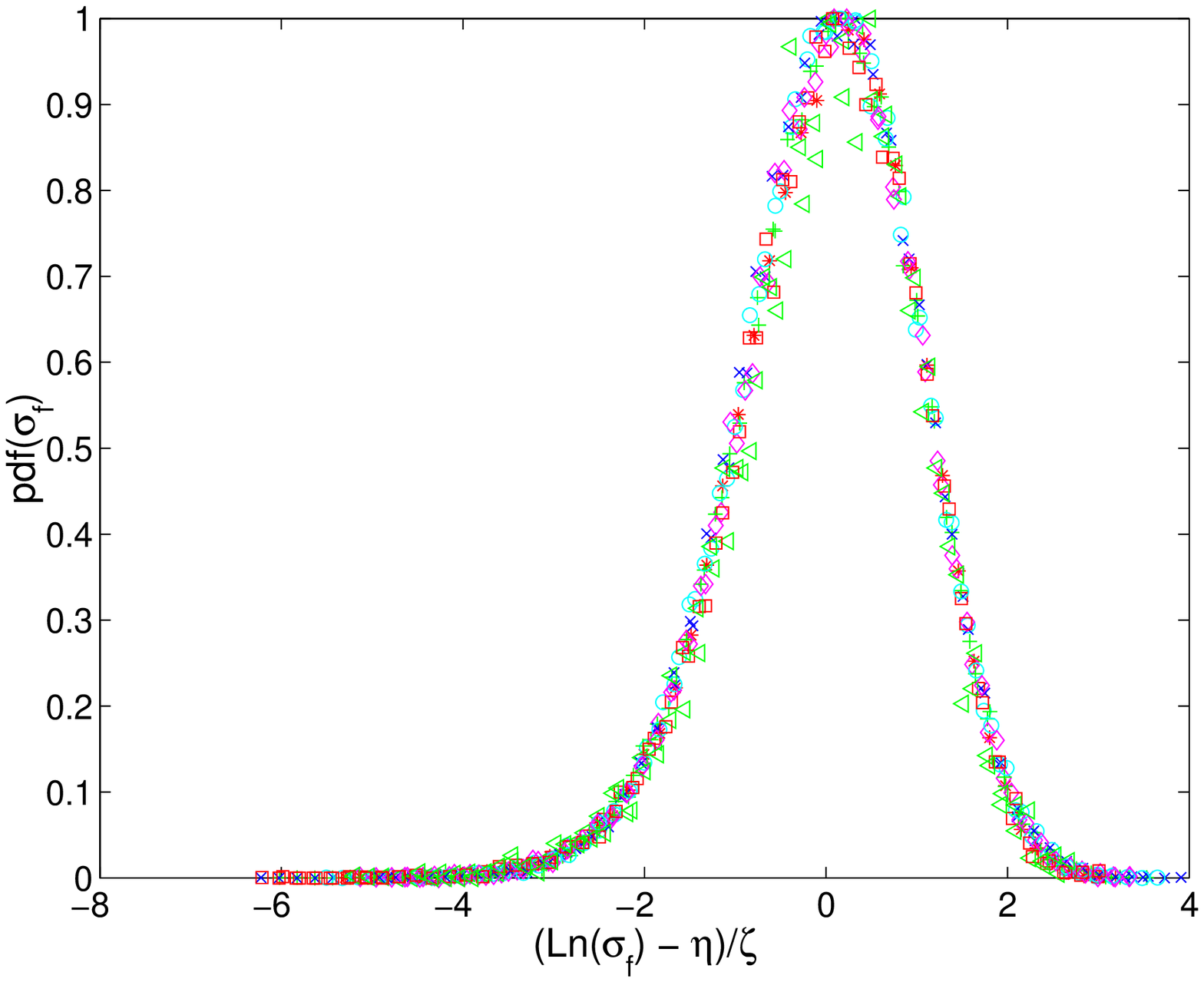}
\includegraphics[width=8cm]{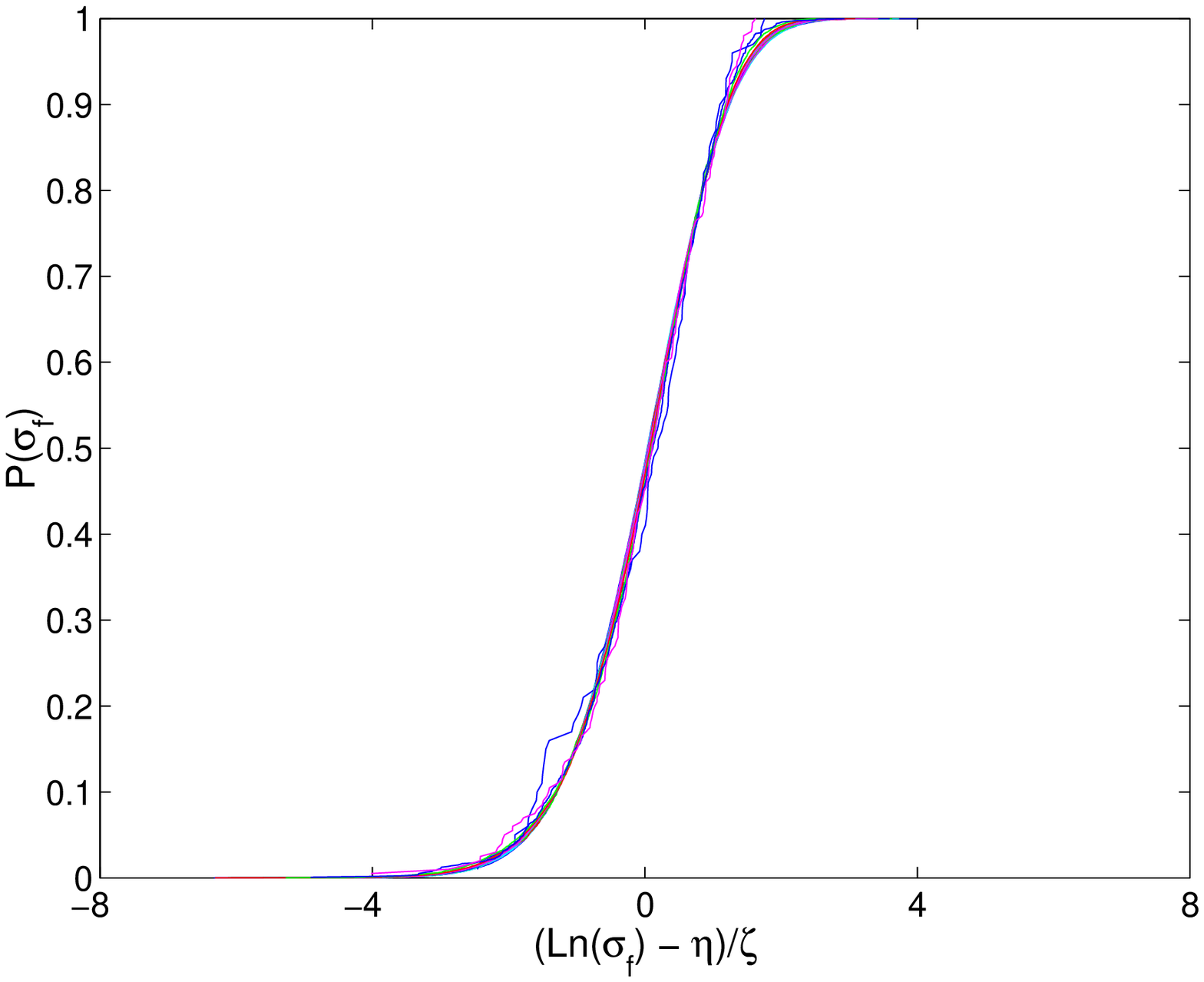}
\caption{Universality of fracture strength distribution in the random
thresholds spring and fuse models.  (a) Fracture strength density
distributions for triangular spring lattices ($L
= \{8,16,24,32,64,128\}$) and triangular fuse lattices 
($L = \{4,8,16,24,32,64,128\}$) with uniform disorder. 
(b) Cumulative fracture strength
distribution for triangular spring lattices ($L =
\{8,16,24,32,64,128,256\}$) and triangular fuse lattices 
($L = \{4,8,16,24,32,64,128,256,512\}$) with uniform disorder. The collapse of
the data in random spring and fuse models suggests universality of
fracture strength distribution. In the RFM case, the universality of 
fracture strength distributions with respect to different 
lattice topologies is presented in Ref. \cite{nukalaepjb}.}
\label{fig:univ_pdf}
\end{figure}

Figure~\ref{fig:univ_pdf}(a) presents the fracture strength density
distributions for random thresholds spring model using the standard
Lognormal variable, $\bar{\xi}$, defined as $\bar{\xi} = \frac{Ln(\sigma_f) -
\eta}{\zeta}$, where $\sigma_f$ refers to the fracture strength
defined as the peak load divided by the system size $L$, and $\eta$
and $\zeta$ refer to the mean and the standard deviation of the
logarithm of $\sigma_f$.  In order to verify the universality of
fracture strength distribution, the fracture strength distributions
from \cite{nukalaepjb} corresponding to random thresholds fuse model
(RFM) using triangular lattice systems with uniform disorder are  
presented in Fig.~\ref{fig:univ_pdf}(a) along with those corresponding to random thresholds spring model.
In particular, Fig.~\ref{fig:univ_pdf}(a) shows the data for different
lattice system sizes, $L$, corresponding to (a) triangular spring
lattice, $L = \{8,16,24,32,64,128\}$ and (b)
triangular fuse lattices of sizes $L =
\{4,8,16,24,32,64,128\}$. In all, there are 13 plots in Fig.~\ref{fig:univ_pdf}(a), and the
excellent collapse of the data for various spring and fuse lattices
clearly indicates the universality of the fracture strength density
distribution. The results presented in Fig.~\ref{fig:univ_pdf}(a) are
limited only up to a system size of $L=128$ due to the availability of
fewer sample configurations for larger lattice systems. In order to
attain a good collapse of the data for the density distributions, it
is necessary to consider many sample configurations. On the other
hand, good collapse of the data for the cumulative distributions can
be achieved using fewer number of sample
configurations. Figure.~\ref{fig:univ_pdf}(b) presents the cumulative
fracture strength versus the standard Lognormal variable, $\bar{\xi}$, for
random spring and fuse lattice networks for system sizes up to $L =
512$.  In particular, in Fig.~\ref{fig:univ_pdf}(b), we plot the
numerical simulation results of RSM for system sizes $L =
\{8,16,24,32,64,128,256\}$ along with those of RFM for system sizes 
$L = \{4,8,16,24,32,64,128,256,512\}$. In all, there are about 16 curves
(7 for triangular spring lattices and 9 for
triangular fuse lattices) in Fig.~\ref{fig:univ_pdf}(b), and the excellent
collapse of the data suggests universality
of fracture strength distribution. In Ref. \cite{nukalaepjb}, we have also presented 
the collapse of the fracture strength distribution for different 
lattice topologies (such as triangular and diamond), which is consistent 
with the notion of universality of fracture strength distribution. That is, $P(\sigma \le \sigma_f) =
\Psi(\xi)$, where $P(\sigma \le \sigma_f)$ refers to the cumulative
probability of fracture strength $\sigma \le \sigma_f$, $\Psi$ is a
universal function such that $0 \le \Psi \le 1$, and $\bar{\xi} =
\frac{Ln(\sigma_f) - \eta}{\zeta}$ is the standard Lognormal variable.

\begin{figure}[hbtp]
\includegraphics[width=8cm]{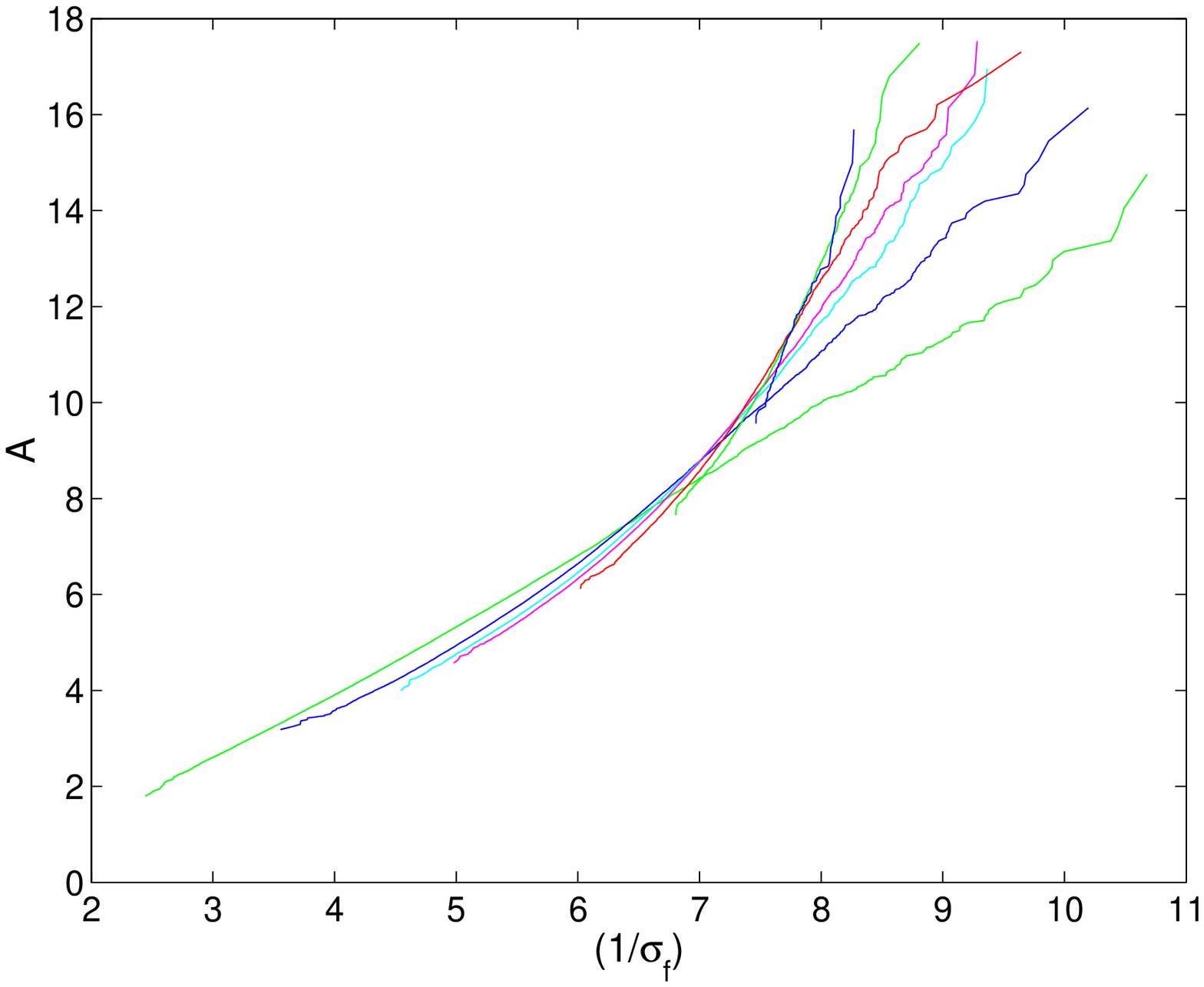}
\includegraphics[width=8cm]{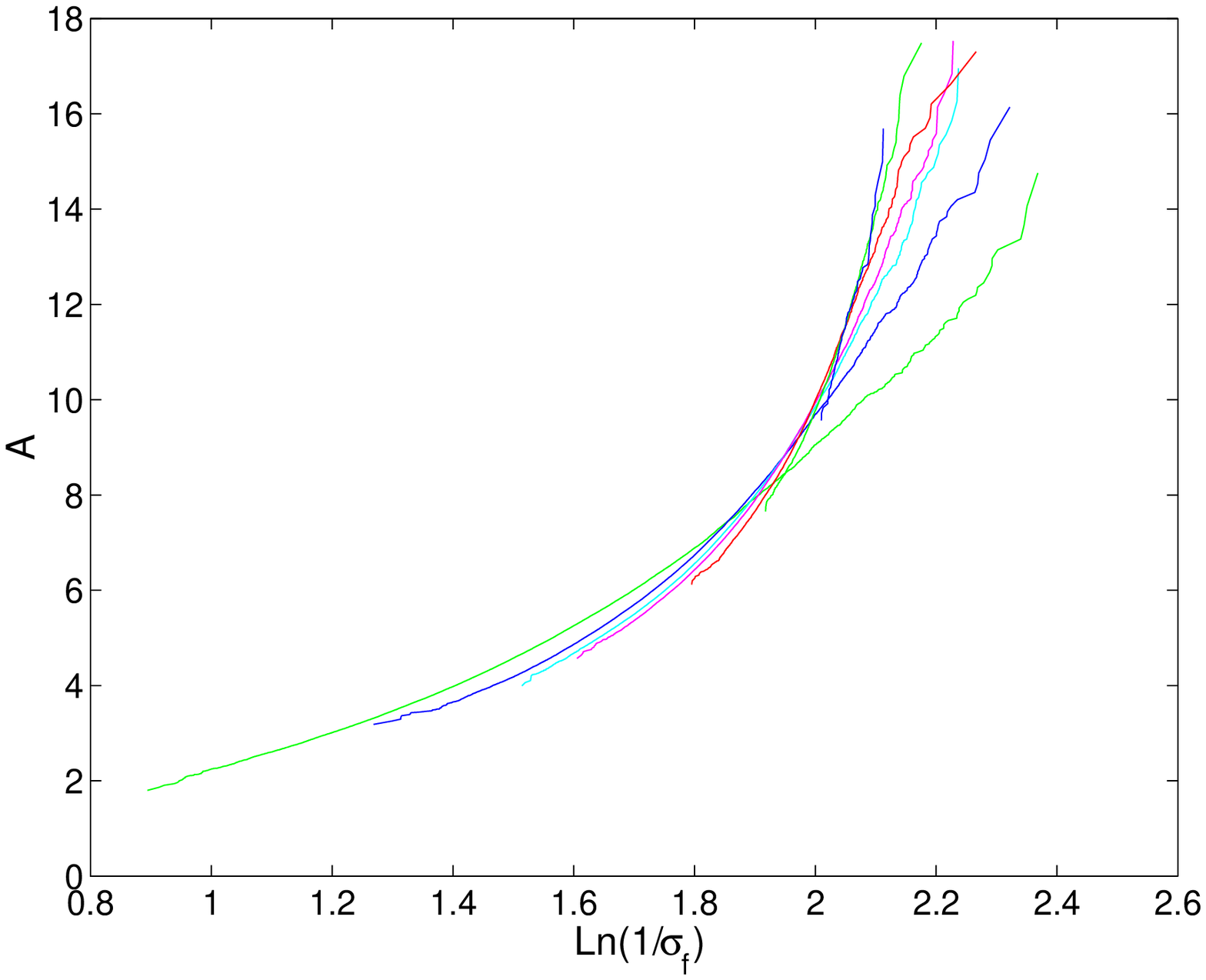}
\includegraphics[width=8cm]{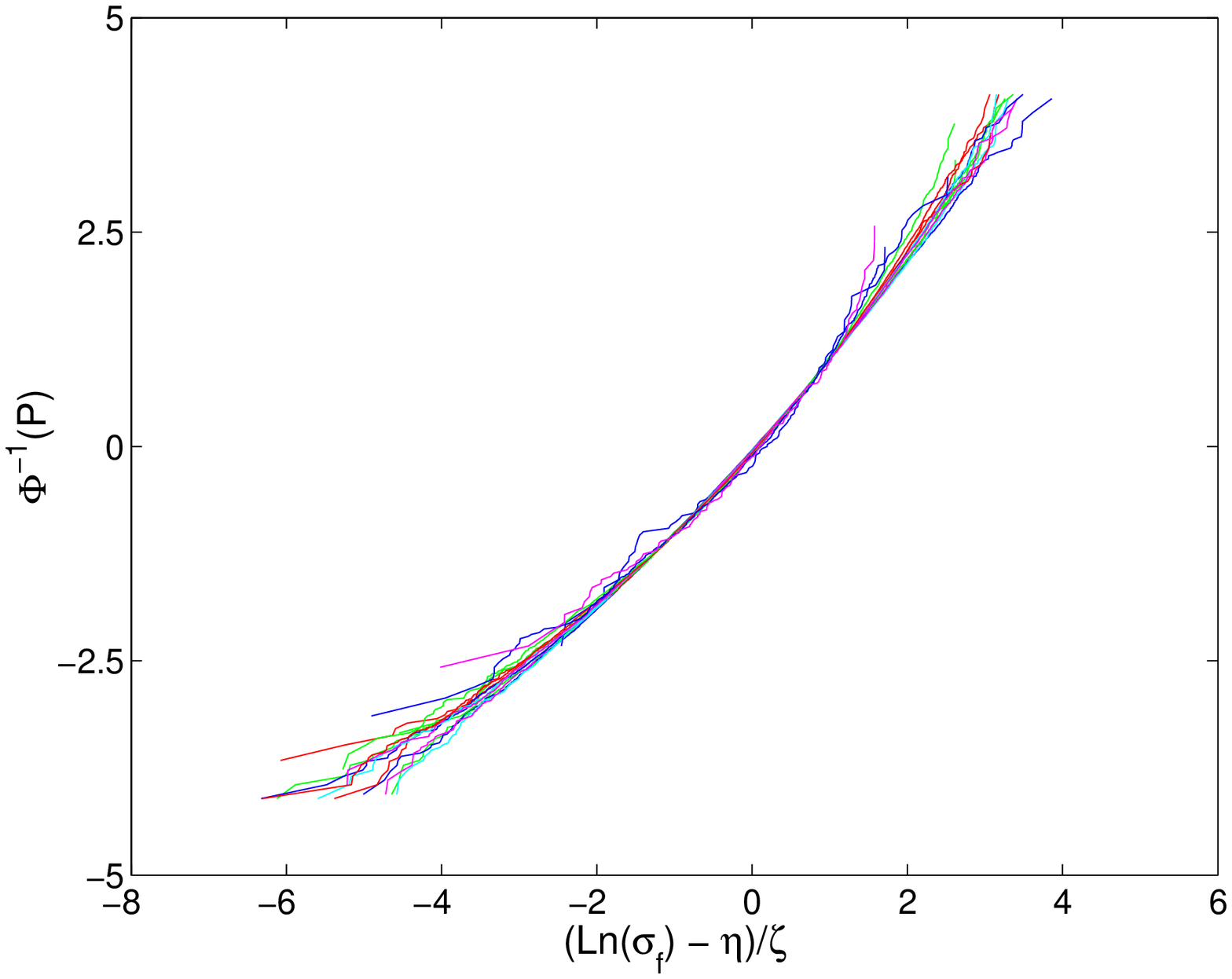}
\caption{Probability distribution fits for fracture strengths at the
peak load in a triangular spring lattice network for different lattice
system sizes $L = \{8,16,24,32,64,128,256\}$. (a) Modified Gumbel
distribution (top). (b) Weibull distribution (middle). (c) Lognormal
distribution fit for all the 16 curves (see
Fig.~\ref{fig:univ_pdf}(b)) (bottom). Since the data for different
lattice system sizes does not collapse onto a single curve, Weibull
distribution may not be an adequate fit for representing fracture
strengths in the RSM. On the other hand, the collapse of the data in
the reparametrized Lognormal distribution fit suggests that the
Lognormal distribution describes the fracture strength distribution
adequately.}
\label{fig:weibull}
\end{figure}

Figures~\ref{fig:weibull}(a) and ~\ref{fig:weibull}(b) present the modified Gumbel and Weibull 
fits for the fracture strength 
distribution of triangular spring lattice network using 
\begin{eqnarray}
A & = & k ~ \left(\frac{1}{\sigma_f^\delta}\right) ~-~ Ln ~c \label{gda}
\end{eqnarray}
for the modified Gumbel distribution, and 
\begin{eqnarray}
A & = & m ~ Ln\left(\frac{1}{\sigma_f}\right) ~-~ Ln ~c \label{wda}
\end{eqnarray}
for the Weibull distribution. In Eqs. \ref{gda} and \ref{wda}, 
$k$, $\delta$, $c$ and $m$ are constants, and $A$ is defined as 
\begin{eqnarray}
A & = & -Ln\left[-\frac{Ln\left(1 - P(\sigma_f)\right)}{L^2}\right] \label{eqnA}
\end{eqnarray}
where $P(\sigma_f)$ denotes the cumulative distribution. 
From these figures, it is 
clear that fracture strength data for different lattice system sizes 
does not collapse on to a single straight line as it should, if the data were to follow
Eq. (\ref{gda}) or (\ref{wda}). This indicates that 
neither modified Gumbel nor Weibull distributions may represent the fracture strengths 
distribution accurately for the RSM. In the Ref. ~\cite{nukalaepjb}, similar conclusion 
has been drawn for the fracture strengths distribution of RFM. 

On the other hand, in Fig.~\ref{fig:weibull}(c), we test the Lognormal
description for fracture strengths by plotting the inverse of the
cumulative probability, $\Phi^{-1}(P(\sigma_f))$, against the standard
Lognormal variable, $\bar{\xi}$. In the above description, $\Phi( \cdot )$
denotes the standard normal probability function.  In particular, in
the Fig.~\ref{fig:weibull}(c) we present the Lognormal fit for the
cumulative fracture strength distributions obtained for random
thresholds spring and fuse models 
(i.e., for all of the 16 curves in Fig.~\ref{fig:univ_pdf}(b)).  
Once again, this figure clearly indicates that
the fracture strength distribution obtained for different lattice
system sizes collapses onto a single curve, albeit minute deviation
from straight line behavior is evident. We have also used the normal
distribution to collapse the fracture strength data of triangular spring and fuse 
lattice systems. Although the data collapse is reasonable, it
is not as good as that of Lognormal distribution based on
Kolmogorov-Smirnov goodness-of-fit test.

\section{Size effects in the mean fracture strength}
\begin{figure}[hbtp]
\includegraphics[width=8cm]{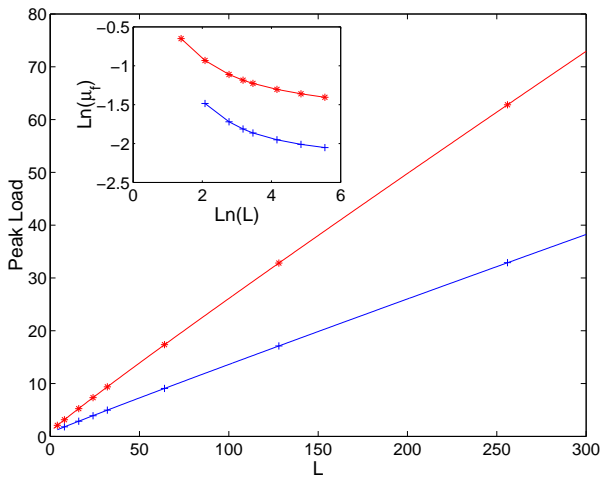}
\caption{Proposed scaling law for the mean fracture strength ($F_{peak} = C_0 L^{\bar{\alpha}} + C_1$). 
(1) Triangular spring network (symbol: -+-): $\bar{\alpha} = 0.97$. 
(2) Triangular fuse network (symbol: -*-): $\bar{\alpha} = 0.956$; 
The corresponding Weibull fit for the mean fracture strength is shown in the inset. 
Nonlinearity of the plots in the inset suggests that mean fracture strength does not 
follow a power law scaling consistent with the Weibull distribution.}
\label{fig:wei_mean}
\end{figure}

The mean fracture strength data for various random threshold spring
lattice system sizes is presented in Table 2. In
Ref.~\cite{nukalaepjb}, for the RFM, we have suggested a scaling form
${\bar{F}}_{peak} = C_0 L^{\bar{\alpha}} + C_1$ for the peak load, where
$C_0$ and $C_1$ are constants.  Correspondingly, the mean fracture
strength defined as $\mu_f = \frac{{\bar{F}}_{peak}}{L}$, is given by
$\mu_f = C_0~L^{\bar{\alpha} - 1} + \frac{C_1}{L}$.  We have used the same
scaling law for the random thresholds spring model as well, and the
result presented in Fig.~\ref{fig:wei_mean} indicates that the
exponent $\bar{\alpha}$ is approximately equal to
$0.97$, which is once again in consistent with the $\bar{\alpha} = 0.96$
obtained for RFM using both triangular and diamond lattice
topologies. The inset in Fig.~\ref{fig:wei_mean} presents a power law
fit $\mu_f \sim L^{-\frac{2}{m}}$ that is consistent with a Weibull
distribution for fracture strengths.  From the nonlinearity of the
plots in the inset of Fig.~\ref{fig:wei_mean}, it is clear that the
mean fracture strength does not follow a simple power law scaling that
is consistent with a Weibull distribution for fracture strengths. We
have also verified that the mean fracture strength does not follow a
scaling law of form $\mu_f^\delta = \frac{1}{A_1 ~+~ B_1 ~Ln ~L}$ that
is consistent with a modified Gumbel distribution for fracture
strengths \cite{duxbury86,duxbury87,duxbury88,beale88}.

Since a very small negative exponent $(\bar{\alpha} - 1)$ is equivalent to a
logarithmic correction, i.e., for $(1-\bar{\alpha}) << 1$, $L^{\bar{\alpha} - 1}
\sim (log(L))^{-\psi}$, an alternative expression for the mean
fracture strength may be obtained as $\mu_f = \frac{\mu_f^\star}{(Log
L)^\psi} + \frac{c}{L}$, where $\mu_f^\star$ and $c$ are constants
that are related to the constants $C_0$ and $C_1$.  This suggests that
the mean fracture strength of the lattice system decreases very slowly
with increasing lattice system size, and scales as $\mu_f \approx
\frac{1}{(Log L)^\psi}$, with $\psi \approx 0.15$, for very large
lattice systems.

\section{Avalanches}

\begin{figure}[hbtp]
\centerline{\psfig{file=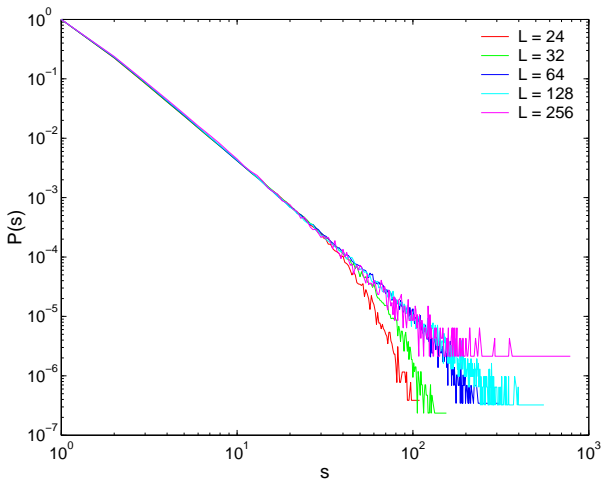,width=8cm,clip=!}}
\caption{The distribution of avalanche sizes (without the last
catastrophic event) for triangular spring lattices of different
sizes.}
\label{fig:avalan}
\end{figure}

\begin{figure}[hbtp]
\centerline{\psfig{file=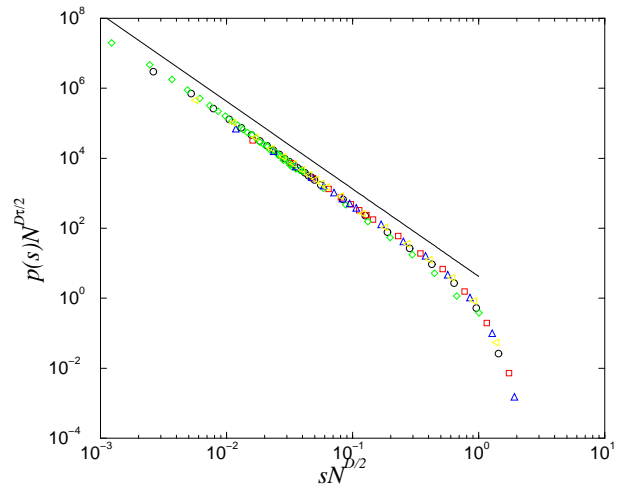,width=8cm,clip=!}}
\caption{Data collapse of the avalanche size distributions excluding
the final catastrophic event. The exponents used for the collapse are
$\tau=2.5$ (the reference line has this slope) and $D=1.1$. The
distributions have been logarithmically binned to reduce
fluctuations.}
\label{fig:avalan_coll}
\end{figure}

The avalanche size distribution, once the last event is excluded, is a
power law followed by an exponential cutoff at large avalanche sizes
(see Fig.~\ref{fig:avalan}) \cite{nota}.  The cutoff size $s_0$ is
increasing with the lattice size, so that we can describe the
distribution by a scaling form
\begin{equation}
P(s,N)=s^{-\tau} g(s/N^{D/2}),
\end{equation}
where $D$ represents the fractal dimension of the avalanches and
$N=(3L+1)(L+1)$ is the number of bonds. Figure \ref{fig:avalan_coll}
presents the data collapse of the distribution of avalanche sizes
using the exponents $\tau=2.5$ and $D=1.1$.

\begin{figure}[hbtp]
\centerline{\psfig{file=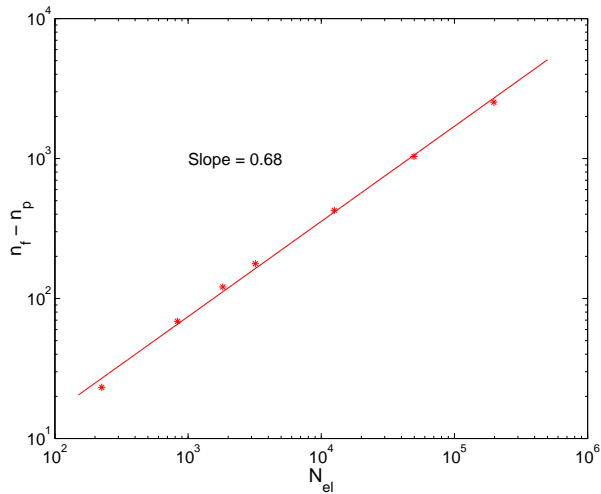,width=8cm,clip=!}}
\caption{The mean avalanche size of the last catastrophic event ($s_m = (n_f-n_p)$) 
scales as a power law of $N_{el}$. Once again, the scaling exponent 
$b=0.68$ for RSM is similar to the scaling exponent $b\simeq 0.7$ obtained for 
RFM using triangular and diamond lattices (see Fig. 14 of Ref. \cite{nukalajstat}).}
\label{fig:4}
\end{figure}

\begin{figure}[hbtp]
\centerline{\psfig{file=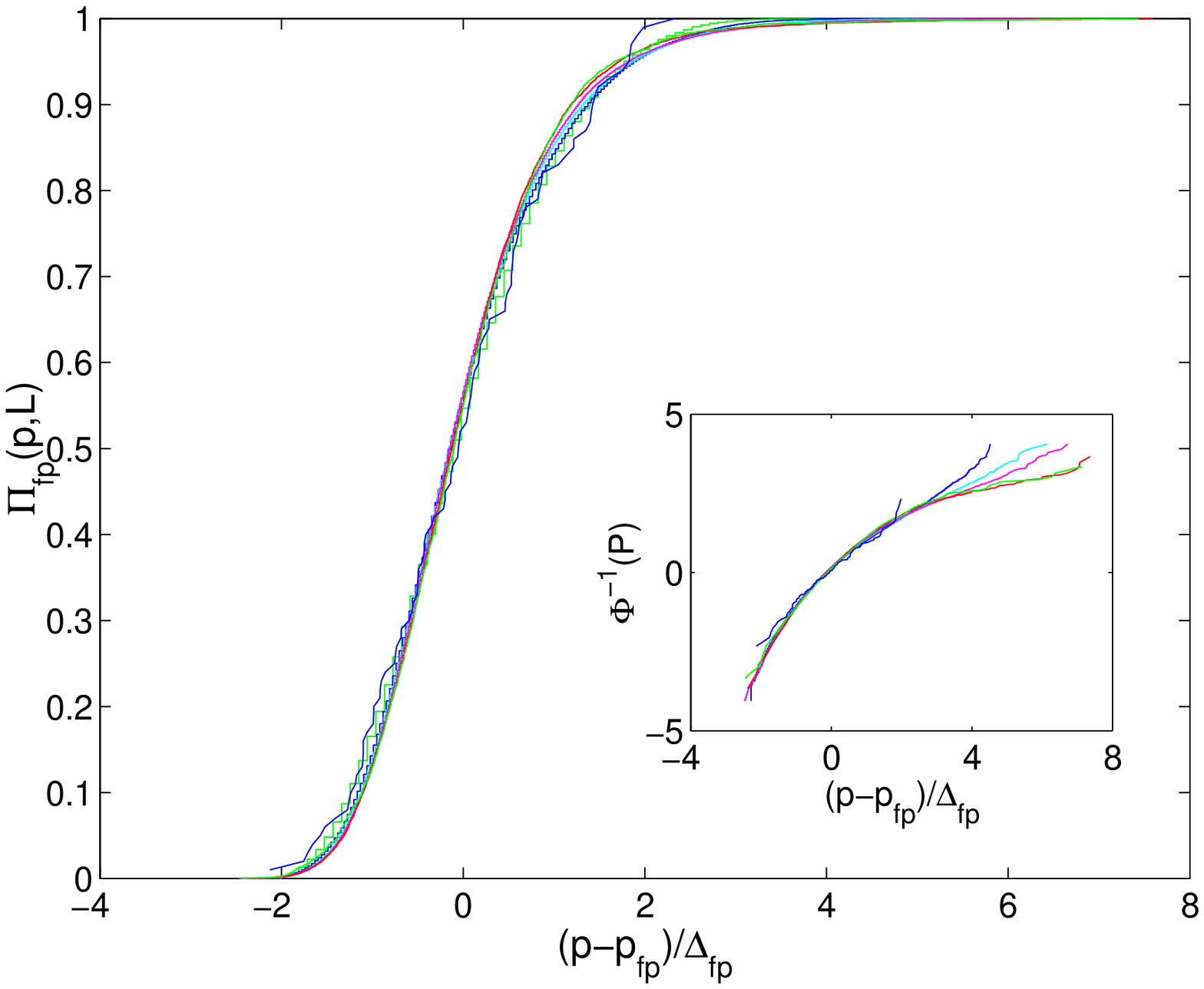,width=8cm,clip=!}}
\centerline{\psfig{file=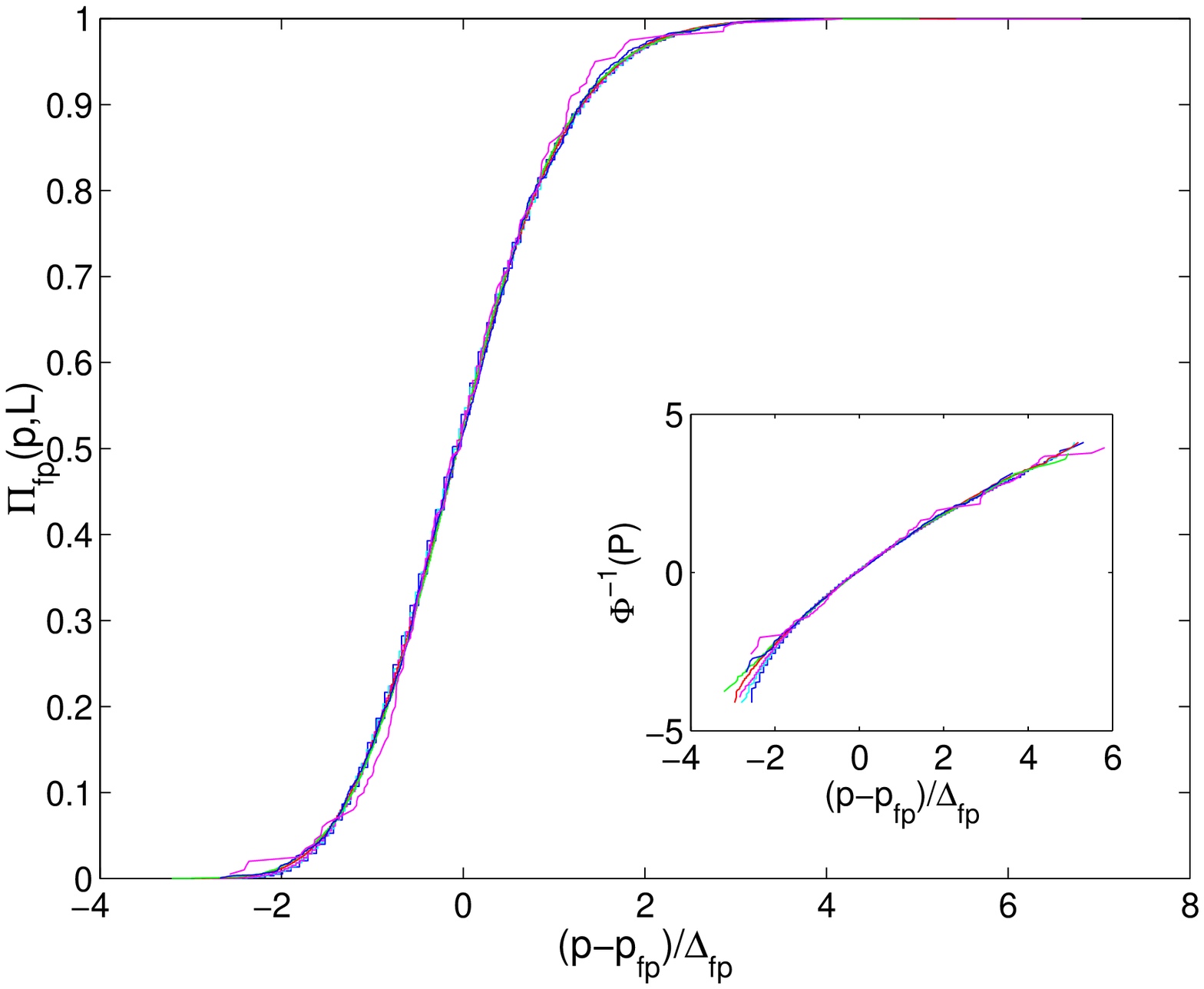,width=8cm,clip=!}}
\caption{The collapsed cumulative distribution of last avalanche. 
(a) RSM (b) RFM. The insets in each of these figures 
show how well the data can be represented by a normal distribution fit. The presence of 
significant nonlinearity of the data in these insets suggests that normal distribution 
may not be an adequate fit for representing the distribution of last 
avalanche size for the RSM model, whereas it may be an adequate fit for the RFM model.}
\label{fig:last_avalan}
\end{figure}

So far we have considered avalanche statistics integrating the
distribution over all the values of the forces upto the peak load, but
the avalanche signal is not stationary: as the force increases so does
the avalanche size. In particular, the last avalanche is much larger
than the others. Its typical size grows as $s_m = (n_f-n_p) \sim N^b$,
with $b\simeq 0.68$ (see Fig. \ref{fig:4}), which is once again in
good agreement with the $b\simeq 0.7$ value obtained for RFM (see
Fig. 14 of Ref.~\cite{nukalajstat}). The cumulative distribution of
last avalanche sizes for the RSM and RFM is presented in
Figs. \ref{fig:last_avalan}(a) and (b) respectively.  While the
distribution is approximately Gaussian for RFM as shown by the data
collapse (almost linear) in the inset of
Fig.~\ref{fig:last_avalan}(b), there appears to be significant
nonlinearity in the data collapse of the plots in the inset of
Fig.~\ref{fig:last_avalan}(a). This suggests that normal distribution
may not be an adequate fit for representing the distribution of last
avalanche size in the RSM model. We notice here that the post-peak  regime
is different in the two models because the RSM can fail because
of a loss in rigidity. In general, the significantly different nature
of the last avalanche with respect to the precursors is revealed
both by the distribution type (Gaussian or power law) and by its
characteristic value, scaling as $2b \simeq 1.36$ or $D=1.1$.  This
difference reflects the fact that the last avalanche is a catastrophic
event corresponding to unstable crack growth, while precursors reflect
metastable crack growth and the two processes are different.

\section{Summary}

In this study we investigated the universality of random thresholds 
spring and fuse models using large
scale numerical simulations and large number of sample
configurations to reduce the statistical error in the numerical
results. For both models, we considered triangular lattice topology 
with uniform disorder and increased the load quasistatically.
We performed several statistical measures characterizing the
fracture process that can be summarized as follows:
\begin{enumerate}

\item {\it Damage localization:} the process of localization is similar in
the RFM and RSM. Damage is accumulated in a uniform manner up to the
peak load and then suddently localizes leading to complete failure.
This process is described by damage profiles, that are basically flat 
until peak load and show a peak, with exponential tails, in the 
post peak regime. The collapse of the damage profiles implies some
small differences in the exponets for the two models.

\item {\it Damage density:} The number of
broken bonds at failure or at peak load scales with the lattice size.
We have shown that the behavior in RFM and RSM is very similar, but in
both cases it is not possibile to distinguish a power law from a linear
behavior with a logarithmic correction.

\item{\it Damage distributions:} The distribution of broken bonds 
at peak load follows the normal distribution for RFM and RSM.

\item {\it Fracture strength:} The fracture strength distribution 
is found to be lognormal for both models and the mean fracture strength
scales logarithmically.

\item {\it Avalanches:} The integrated avalanche distributions follow
a power law in both models. The results of the RSM yields an exponent $\tau=5/2$ that is 
very close to the exponent observed in global load sharing fiber bundle model (FBM),
while larger deviations are found
in the RFM (i.e. $\tau=2.7$ \cite{ZAP-05}).

\end{enumerate}

Thus in conclusion, we can state that RFM and RSM are qualitatively
very similar: distributions have the same forms, localization proceeds
in the same way, avalanches are similar. The only diffirences can be
found in small quantitative deviations in exponents. We can not rule
out that these are due to differences in the finite size behavior of
the models and that at large scales the behavior is the same. In addition,
the rigidity mechanism present in the RSM and not in the RSM could 
explain some deviations in the post-peak regime.
 
\par
\vskip 1.00em%
\noindent
{\bf Acknowledgment} \\ This research is sponsored by the
Mathematical, Information and Computational Sciences Division, Office
of Advanced Scientific Computing Research, U.S. Department of Energy
under contract number DE-AC05-00OR22725 with UT-Battelle, LLC.

\begin{table}[hbtp]
  \leavevmode
  \begin{center}
  \vspace*{1ex}
  \begin{tabular}{|c|c|c|c|c|c|}\hline
  L  & $N_{config}$ & \multicolumn{4}{c|}{Triangular} \\\cline{3-6}
     &  & $p_p$ & $\Delta_p$ & $p_f$ & $\Delta_f$ \\
  \hline
  8 & 40000  & 0.1213 & 0.0285 & 0.2244 & 0.0482 \\
 16 & 40000  & 0.1045 & 0.0179 & 0.1869 & 0.0349 \\
 24 & 40000  & 0.0970 & 0.0137 & 0.1633 & 0.0258 \\
 32 & 40000  & 0.0923 & 0.0113 & 0.1477 & 0.0201 \\
 64 & 8000  & 0.0835 & 0.0075 & 0.1175 & 0.0106 \\
128 & 2400  & 0.0763 & 0.0051 & 0.0972 & 0.0056 \\
256 & 100  & 0.0708 & 0.0031 & 0.0836 & 0.0029 \\
  \hline
  \end{tabular}
  \label{table1}
  \end{center}\caption{Mean and standard deviation of damage density at the peak load 
and failure in the random thresholds spring model using triangular lattice network 
with uniform disorder distribution. $N_{config}$ denotes the number of 
configurations used in averaging the results for each system size. $p_p$ and $p_f$ denote 
the mean fraction of broken bonds in a lattice system of size $L$ at the peak load and 
at failure, respectively. Similarly, $\Delta_p$ and $\Delta_f$ denote the standard deviation 
of the fraction of broken bonds at the peak load and at failure respectively.}
\end{table}

\begin{table}[hbtp]
  \leavevmode
  \begin{center}
  \vspace*{1ex}
  \begin{tabular}{|c|c|c|c|}\hline
  L  & $N_{config}$ & \multicolumn{2}{c|}{Triangular} \\\cline{3-4}
     &  & Mean & Std \\
  \hline
  8 & 40000  & 1.8125 & 0.3318 \\
 16 & 40000  & 2.8646 & 0.3364 \\
 24 & 40000  & 3.9170 & 0.3558 \\
 32 & 40000  & 4.9619 & 0.3761 \\
 64 & 8000  & 9.0865 & 0.4632 \\
128 & 2400  & 17.1286 & 0.6122 \\
256 & 100  & 32.8959 & 0.8024 \\
  \hline
  \end{tabular}
  \label{table2}
  \end{center}\caption{Peak load in the random thresholds spring model using triangular 
lattice network with uniform disorder distribution. $N_{config}$ denotes the number of 
configurations used in averaging the results for each system size.}
\end{table}


\begin{thebibliography}{99}
\bibitem{nota}
Removing the last event implies that the peak observed 
in Ref.~\protect\cite{politi} does not appear here. See also
Ref.~\protect\cite{ZAP-05}.
\end{thebibliography}
\end{document}